\documentclass{aa}  

\usepackage{graphicx}
\usepackage{txfonts}
\usepackage{float}
\begin{document}

   \title{The 492 GHz emission of Sgr A* constrained by ALMA}


   \author{Hauyu Baobab Liu\inst{1}\fnmsep\inst{2}
          \and
          Melvyn C. H. Wright\inst{3}\fnmsep 
          \and
          Jun-Hui Zhao \inst{4}
          \and 
          Elisabeth A. C. Mills \inst{5}
          \and
           Miguel A. Requena-Torres\inst{6}
          \and 
          Satoki Matsushita \inst{2}
          \and
          Sergio Mart\'in \inst{7}\fnmsep\inst{8}
          \and 
          J\"{u}rgen Ott \inst{5}
          \and 
          Mark R. Morris \inst{9}
          \and 
          Steven N. Longmore \inst{10}
          \and 
          Christiaan D. Brinkerink \inst{11}
          \and 
          Heino Falcke \inst{11}
          }

   \institute{European Southern Observatory (ESO), Karl-Schwarzschild-Str. 2, D-85748 Garching, Germany \\
                 \email{baobabyoo@gmail.com}
         \and
             Academia Sinica Institute of Astronomy and Astrophysics, P.O. Box 23-141, Taipei, 106 Taiwan 
         \and
            Department of Astronomy, Campbell Hall, UC Berkeley, Berkeley, CA 94720
         \and
             Harvard-Smithsonian Center for Astrophysics, 60 Garden St, MS 78, Cambridge, MA 02138
         \and 
         National Radio Astronomy Observatory, 1003 Lopezville Rd, Socorro, NM 87801, USA
         \and 
         	Space Telescope Science Institute, 3700 San Martin Dr., Baltimore,MD 21218, USA
         \and
            European Southern Observatory, Alonso de C\'ordova 3107, Vitacura, Santiago
         \and
            Joint ALMA Observatory, Alonso de C\'ordova 3107, Vitacura, Santiago, Chile
         \and 
         Dept. of Physics \& Astronomy, University of California, Los Angeles, CA 90095-1547  USA
         \and 
            Astrophysics Research Institute, Liverpool John Moores University, 146 Brownlow Hill, L3 5RF
         \and 
            Department of Astrophysics/IMAPP Radboud University Nijmegen P.O. Box 9010 6500 GL Nijmegen The Netherlands
             }

   \date{Received February 15, 2016; accepted March XX, 2016}



  \abstract
   {}
   {Our aim is to characterize the polarized continuum emission properties including intensity, polarization position angle, and polarization percentage  of Sgr A* at $\sim$492 GHz. This frequency being well into the submillimeter-hump where the emission is supposed to become optically thin, allows us to see down to the event horizon. Hence the reported observations contain potentially vital information on black hole properties. We have compared our measurements with previous, lower frequency observations, which provides information in the time domain.}
   {We report continuum emission properties of Sgr A* at $\sim$492 GHz, based on the Atacama Large Millimeter Array (ALMA) observations. We measured fluxes of Sgr A* from the central fields of our ALMA mosaic observations. We used the observations of the likely unpolarized continuum emission of Titan, and the observations of C\textsc{i} line emission, to gauge the degree of spurious polarization.}
   {The flux of 3.6$\pm$0.72 Jy during our run is consistent with extrapolations from the previous, lower frequency observations. We found that the continuum emission of Sgr A* at $\sim$492 GHz shows large amplitude differences between the XX and the YY correlations. The observed intensity ratio between the XX and YY correlations as a function of parallactic angle may be explained by a constant polarization position angle of $\sim$158$^{\circ}$$\pm$3$^{\circ}$. The fitted  polarization percentage of Sgr A* during our observational period is 14\%$\pm$1.2\%. The calibrator quasar J1744-3116 we observed at the same night can be fitted to Stokes I = 252 mJy, with 7.9\%$\pm$0.9\% polarization in position angle P.A. = 14.1$^{\circ}$$\pm$4.2$^{\circ}$.} 
   {The observed polarization percentage and polarization position angle in the present work appear consistent with those expected from longer wavelength observations in the period of 1999-2005. In particular, the polarization position angle at 492 GHz, expected from the previously fitted 167$^{\circ}$$\pm$7$^{\circ}$ intrinsic polarization position angle and (-5.6$\pm$0.7)$\times$10$^{5}$ rotation measure, is 155$^{+9}_{-8}$, which is consistent with our new measurement of polarization position angle within 1$\sigma$. The polarization percentage and the polarization position angle may be varying over the period of our ALMA 12m Array observations, which demands further investigation with future polarization observations.}

   \keywords{black hole physics --- Galaxy: center --- polarization --- submillimeter --- techniques: interferometric
               }

   \maketitle
%

\section{Introduction}
The sub-Eddington accretion of the nearest supermassive black hole, Sgr A* ($\sim$4$\times$10$^{6}$ $M_{\odot}$, e.g. Sch{\"o}del et al. 2002; Ghez et al. 2005; Gillessen et al. 2009), has inspired a tremendous amount of observational and theoretical activity (see Yuan \& Narayan 2014 for a complete review of existing theories).
This {\bf includes} monitoring observations at multiple wavelengths to probe synchrotron emission, which may be from the innermost part of an accretion flow, or the footpoint of a jet (Falcke et al. 2000; Liu et al. 2007; Falcke et al. 2009; Huang et al. 2009; more below), and has motivated very long baseline millimeter interferometric observations (e.g. Johnson et al. 2015, and references therein).

\begin{table*}
\caption{Observed fluxes of Sgr A* (from a vector averaging at Sgr A* in the visibility domain, prior to correction of primary beam attenuation)}
\label{tab:obs}
\hspace{0cm}
\hspace{1.7cm}
\begin{tabular}{ p{1.5cm} cccc } \hline\hline
Field ID		&		Correlation		&		Average amplitude		&		Amplitude Standard Deviations		&		Parallactic angle \\
				&							&		(Jy)							&		(Jy)						&		($^{\circ}$)				\\\hline
18				&		XX				&		2.23							&		0.54						&		-42.4						\\
				&		YY				&		1.85							&		0.52						&									\\
25				&		XX				&		2.05							&		0.54						&		-40.3						\\
				&		YY				&		1.69							&		0.51						&									\\
0				&		XX				&		4.18							&		0.54						&		-23.0						\\
				&		YY				&		3.14							&		0.53						&									\\
94				&		XX				&		2.39							&		0.69						&		5.5						\\
				&		YY				&		2.03							&		0.64						&									\\
101			&		XX				&		1.99							&		0.66						&		9.0						\\
				&		YY				&		1.76							&		0.63						&									\\
133			&		XX				&		1.83							&		0.62						&		28.6						\\
				&		YY				&		2.00							&		0.64						&									\\
134			&		XX				&		1.75							&		0.62						&		29.0						\\
				&		YY				&		1.89							&		0.63						&									\\\hline
\end{tabular}\par
\footnotesize{{\bf $^1$These measurements were taken from the 12m-Array observations.}
$^2$The apparently higher amplitudes measured from field 0 than from the other fields is because that Sgr A* was observed approximately at the center of field 0 (see Figure \ref{fig:fields}).}
\vspace{0.1cm}
\end{table*}

Observations of the polarization position angle and the polarization percentage of the synchrotron emission over a broad range of frequency, may provide information about the geometry and the magnetic field configuration of the accretion flow (Bromley et al. 2001; Liu et al. 2007; Huang et al. 2009), and can diagnose the black hole accretion rate on small {\bf scales} via deriving Faraday rotation (more below).
Previous strong constraints on the linear polarization percentage in the 4.8-112 GHz bands (Bower et al. 1999a, 1999c, 2001), and the detected linear polarization at the 83-400 GHz bands (Aitken et al. 2000; Bower et al. 2003, 2005; Macquart et al. 2006; Marrone et al. 2006a, 2007), have given rise to a model in which linearly polarized radiation is emitted from within a few gravitational radii around Sgr A*, and is further Faraday depolarized by the ionized accretion flow foreground to Sgr A*.
This model is supported by the detection of circularly polarized emission in the 1.4-15 GHz bands (Bower et al. 1999b; Bower et al. 2002; Sault \& Macquart 1999; see also the measurements at 230 and 345 GHz by Mu{\~n}oz et al. 2012).
These observations have constrained the accretion rate of Sgr A* to be between 2$\times$10$^{-9}$ and 2$\times$10$^{-7}$ $M_{\odot}$yr$^{-1}$.
On the other hand, the observed variation of Sgr A*, including large millimeter flares (Zhao et al. 2003, 2004; Marrone et al. 2006), indicates that the accretion may not be stationary. 

In this work, we report new constraints on the polarized emission of Sgr A* at 492 GHz, based on Atacama Large Millimeter Array (ALMA) 12m-Array and Compact Array (ACA) mosaic observations towards the Galactic center.
Our new high-frequency observations provide important, long lever arms in the frequency and time domains for comparison with submillimeter, millimeter, and radio bands observations carried out between 1999 and 2005.
In particular, our observing frequency should be above the turnover frequency at which the emission becomes optically thin (Marrone et al. 2006b). 
Moreover, we are able to reliably diagnose polarizion, which provides the highest frequency interferometric polarization observations so far, and hence tells of the intrinsic polarization.
Our works are pioneering future observations to probe variability, which are crucial to understand the physics of Sgr A*.

Details of our observations and data reduction are provided in Section \ref{chap_obs}.
Our results are given in Section \ref{chap_result}.
In Section \ref{chap_discussion} we address potential systematic biases, and present the comparison of our results with previous observations.
A brief conclusion is provided in Section \ref{chap_conclusion}.

\section{Observations and Data Reduction} 
\label{chap_obs}

The ALMA 12m-Array (consisting of 12 m dishes) mosaic observations of 149 fields were carried out on 2015 April 30 (UTC 06:48:32.4--08:04:38.4), with 39 antennas.
The array consisted of 19 Alcatel antennas (DA), 18 Vertex antennas (DV), and 2 Mitsubishi antennas (PM).
These observations approximately covered a 55$''$$\times$80$''$ rectangular region.
The pointing and phase referencing center of the central field was R.A. (J2000) =17$^{\mbox{\scriptsize{h}}}$45$^{\mbox{\scriptsize{m}}}$40$^{\mbox{\scriptsize{s}}}$.036, and decl. (J2000) =-29$^{\circ}$00$'$28$''$.17, which is approximately centered upon Sgr A*.
We configured the correlator to provide four 1.875 GHz wide spectral windows (spws), covering the frequency ranges of 491.3-493.2 GHz (spw 0), 489.3-491.2 GHz (spw 1), 479.2-481.1 GHz (spw 2), and 481.0-482.9 GHz (spw 3), respectively. 
The observations were designed to cover the C\textsc{i} line and the CS 10-9 line, with rest frequencies are 492.16065 GHz and 489.75093 GHz, respectively.
The frequency channel spacing was 1953.125 kHz ($\sim$1.2 km\,s$^{-1}$).
The receivers are aligned in a parallel-linear configuration, which yielded the XX and YY linear correlations.
The X polarization of the receivers is aligned radially in the receiver cryostat, with Y being aligned perpendicular to X (private communications with Ted Huang and Shin'ichiro Asayama).  
According to ALMA specifications, the accuracy of this alignment is within 2 degrees.
The absolute feed alignment was obtained from the raw data, using the CASA software package (McMullin et al. 2007) command {\tt tb.getcol('RECEPTOR\_ANGLE')}, and can be referenced from the ALMA Cycle 3 {\bf and Cycle 4} Technical Handbook\footnote{https://almascience.eso.org/proposing/call-for-proposals/technical-handbook}.

The range of {\it uv} spatial frequencies sampled by the 12m-Array observations is 25-570 k$\lambda$.
The system temperature ($T_{sys}$) ranged from $\sim$500-1000 K.
The mosaic field was Nyquist sampled in hexagonal packing, with an on-source integration time of 12.08 seconds for each of the 149 mosaic fields.
We observed J1744-3116 approximately every 10 minutes for gain calibrations.
We observed Titan and J1833-2103 for absolute flux and passband calibrations, respectively.

The Atacama Compact Array (ACA; consisting of ten 7 m dishes) observations were carried out on 2015 April 30 (UTC 05:35:00.0--07:30:00.1) with 10 available antennas.
All 10 antennas shared an identical (Mitsubishi, 7m) design. 
The ACA observations approximately covered the same field of view as the 12m-Array mosaic.
The pointing and phase referencing center of the central field was also on Sgr A*.
The correlator setup of the ACA observations was identical to that of the 12m-Array mosaic. 
The ACA observations sampled a {\it uv} spacing range of 14-80 k$\lambda$.
The mosaic field was Nyquist sampled in hexagonal packing.
Due to unspecified technical issues, the ACA observations were terminated at the middle of the track.
Therefore, the southeastern half of the observed region had a 60.6 seconds on-source integration time for each mosaic field, while the northwestern half had a 30.3 seconds on-source integration time for each mosaic field.
This led to different sensitivity and $uv$ coverages  for the southeastern and the northwestern fields. 
Like the 12 m observations, $T_{sys}$ values ranged from $\sim$500-1000 K.
We again observed J1744-3116 approximately every 10 minutes for gain calibrations, and observed Titan and J1517-2422 for absolute flux and passband calibrations, respectively.

There are currently no available single-dish data to provide information on the zero-spacing fluxes for these observations.

A priori calibrations including the application of $T_{sys}$ data, the water vapor radiometer (wvr) solution (which is only provided for the 12m-Array observations), antenna based passband calibrations, gain amplitude and phase calibrations, and absolute flux scaling, were carried out using the CASA software package (McMullin et al. 2007) version 4.3.1.
To enhance the signal to noise ratio, we first solved for and applied phase offsets between the four spectral windows, based on scans on the passband calibrator. We then derived gain calibration solutions. The gain phase solutions were derived separately for the XX and YY correlations,
while the gain amplitude solutions were derived from the average of XX and YY correlations. We derived gain phase solutions for both individual spectral windows and averaging all spectral windows together. We ultimately chose to use the latter, as the wvr solutions for the 12m-Array data in spw 1 and 3 have poorer qualities, which led to massive data flagging when deriving gain phase solutions for individual spectral windows independently. We also tested whether applying or not applying the wvr solutions changed the quality of our final images; ultimately although the difference was minimal, we chose to apply the wvr solutions to the 12m data.
We confirmed that the qualities of continuum images generated from all spectral windows are consistent (e.g., any differences are a result of the available bandwidths in spectral line-free channels).
There was also significant interference due to atmospheric lines in spw 3, which degraded its continuum sensitivity.

The absolute flux scaling was derived incrementally from the gain amplitude solutions, combining all scans.
The scans on Titan were largely flagged due to interference from spectral lines.
Therefore, absolute flux referencing for both the 12m-Array and ACA observations is subject to a large uncertainty (e.g. $\sim$20 \%, empirically). 
This can lead to the mismatched flux levels between the 12m-Array and the ACA observations, and errors in the observed spectral indices.

We fitted the continuum baselines from line-free channels, using the CASA task {\tt uvcontsub}.
After executing {\tt uvcontsub}, we generated a continuum data set for each spectral window, by averaging the line free channels. 
We then exported the calibrated continuum data and the continuum-subtracted line data in standard fits format files, using the CASA task {\tt exportfits}.
Finally, we used the Miriad 4.3.8 (Sault et al. 1995) task {\tt fits} to convert the fits format data into the Miriad data format, for further analyses including imaging. 

We then used Miriad to make synthesized images (i.e. dirty images) of the continuum using naturally weighed data for the 12m-Array and ACA with beam widths (FWHM) $\theta_{\mbox{\scriptsize{maj}}}$$\times$$\theta_{\mbox{\scriptsize{min}}}$ = $0\farcs70$$\times$$0\farcs42$ (P.A.=-88$^{\circ}$) and $\theta_{\mbox{\scriptsize{maj}}}$$\times$$\theta_{\mbox{\scriptsize{min}}}$ = $3\farcs4$$\times$$2\farcs2$ (P.A.=78$^{\circ}$) , respectively. For the C\textsc{i} line, we tapered the 12m-Array data using a Gaussian weighting function of FWHM = $1\farcs5$ to enhance the signal-to-noise ratio of the line, and then generated the  synthesized images. For all of these images we do not make deconvolved C\textsc{i} line (i.e. {\tt clean}ed) maps, to avoid any possibility of uncertainties caused by the {\tt clean} process. 

\begin{figure}
\vspace{-2.0cm}
\hspace{-0.8cm}
\hspace{0.15cm}
\begin{tabular}{p{7.5cm} }
\includegraphics[width=10.5cm]{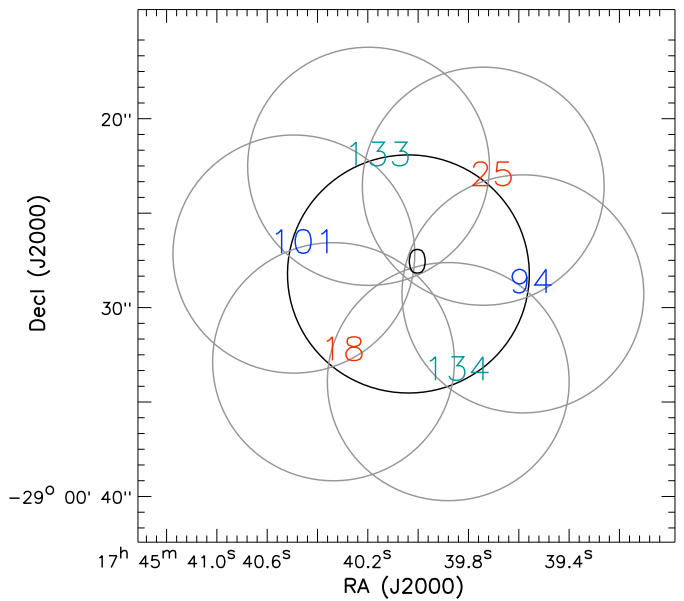} \\
\end{tabular}
\caption{\small{The central 7 mosaic field of views of the 12m-Array observations. 
Black circle shows the field (FWHM) centered on Sgr A*.
Gray circles show the 6 fields which are the nearest to the central one. 
The diameter of these circles is $12\farcs6$.
IDs of the fields which are covered in the same target source scan (i.e. a target source scan is defined by a time period bracketed by two scans on gain calibrator), are labeled with the same color.
These fields were observed in a time sequence of 18$\rightarrow$25$\rightarrow$0$\rightarrow$94$\rightarrow$101$\rightarrow$133$\rightarrow$134 (see also Figure \ref{fig:xxyytime}).
}}
\label{fig:fields}
\vspace{-0.1cm}
\end{figure}

\section{Results}
\label{chap_result}

Throughout this manuscript, the X and Y polarization, and Stokes Q, are defined in the receiver coordinate frame if not specifically mentioned.
In the nearly ideal observational and instrumental condition, the polarization percentage and the polarization position angle of a non-variable source are related to these quantities based on the following formula:
\begin{equation}
\frac{Q}{I} - \delta \equiv \frac{XX- YY}{2I} - \delta = P\cdot\cos(2(\Psi - \eta - \phi) ),
\end{equation}
where $Q$ denotes the observed Stokes Q flux, $\delta$ (Q offset, hereafter) is an assumed constant normalized offset of observed Stokes Q due to amplitude calibration errors or polarization leakage; $P$ is the polarization percentage; $\Psi$, $\eta$, and $\phi$ are the polarization position angle in the sky (e.g. right ascension/declination) frame, the parallactic angle, and the angular separations between the X polarization and the local vertical (which is known as {\it Evector}).
Evector of ALMA is 0$^{\circ}$ for the frequency band we observed.
A wide coverage of $\eta$ during the observations, will allow unambiguously fitting $\delta$, $P$ and $\Psi$.

\subsection{Continuum data}
After a priori calibrations, we found that the continuum emission from Sgr A* was significantly detected in the central 19 mosaic fields {\bf of the 12m-Array observations}.
To inspect the residual phase errors, we used the CASA task {\tt fixvis} to shift the phase referencing centers of these fields to the position of Sgr A*.
We observed up to $\sim\pm$50$^{\circ}$ of residual phase offsets, and a phase RMS of $\sim$16.5$^{\circ}$.
The phase offsets and phase RMS of the XX and YY correlations are consistent.

\begin{figure}
\hspace{-1.5cm}
\hspace{0.15cm}
\begin{tabular}{p{7.5cm} }
\includegraphics[width=10.5cm]{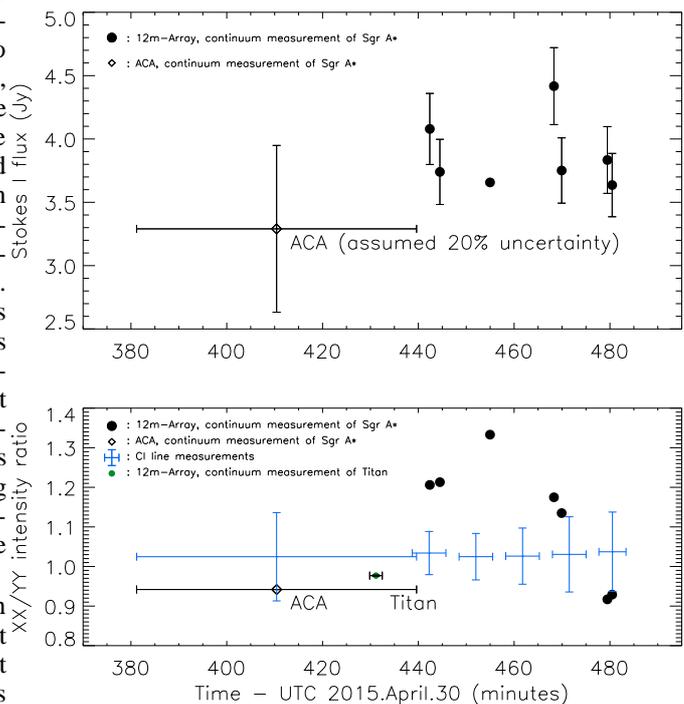} \\
\end{tabular}

\vspace{-0.5cm}
\caption{\small{
{\it Top :--} Fluxes of Sgr A*, measured from images made using only the inner 7 12m-Array mosaic fields, and the flux of Sgr A* from an average of the ACA observations, are both plotted against time. 
Error bars in the horizontal direction represent the scan duration. 
Vertical error bars in the upper panel include both the uncertainty on the pointing ( $\sim$1$''$) and the uncertainty on the primary beam response function. 
The vertical error bar of the ACA data additionally includes a potential 20\% absolute flux calibration uncertainty, relative to the 12m-Array observations.
Horizontal error bars for the continuum data of the 12m-Array observations are shorter than the symbol size.
{\it Bottom :--} The intensity ratio between the XX and the YY intensity maps derived from the continuum and the C\textsc{i} line observations are plotted against time.
Filled black and green symbols show the 12m-Array measurements from Sgr A* and Titan, respectively.
Errors are negligibly small for the snapshot on field 0, which is centered on Sgr A* (see Figure \ref{fig:fields}). 
The C\textsc{i} line measurements are averages from high S/N spectral channels (see also Figure \ref{fig:xxyyfreq}).
Their vertical error bars are given by $\pm$1 standard deviations of the intensity ratio, which were derived from those high S/N spectral channels.
}}
\label{fig:xxyytime}
\vspace{-0.1cm}
\end{figure}

We attribute the phase errors to phase variations that are faster than our gain calibration cycle time, as well as phase offsets between the gain calibrator and the target source fields. 
To correct for these phase errors we used the Miriad task {\tt demos}, assuming the nominal ALMA primary beam shape, to generate models of Sgr A* for the central 7 mosaic fields (Figure \ref{fig:fields}).
We removed the phase errors of the central 7 fields using the Miriad task {\tt selfcal options=mosaic}, with a 0.01 minute solution interval.
Then, we used the Miriad task {\tt uvflux} to fit the observed amplitudes from the visibility data.
Our {\bf 12m-Array} measurements for Sgr A* are summarized in Table \ref{tab:obs} and Figure \ref{fig:xxyytime}.
After self-calibration, the averaged flux of Sgr A* at 492 GHz is 3.6$\pm$0.72 Jy.
The application of phase self-calibration solutions does not significantly change the observed amplitude (or flux) ratios between the XX and the YY correlations.
We do not present flux measurements of Sgr A* from outside of the central 7 mosaic fields due to the potential for large amplitude uncertainties induced by antenna pointing errors (e.g. up to $\sim$1$''$, according to private communication among members in the ALMA Regional Centers), and the poorly understood primary beam phase responses.

\begin{figure*}
\hspace{0.15cm}
\begin{tabular}{p{8cm} p{8cm}}
\includegraphics[width=9.5cm]{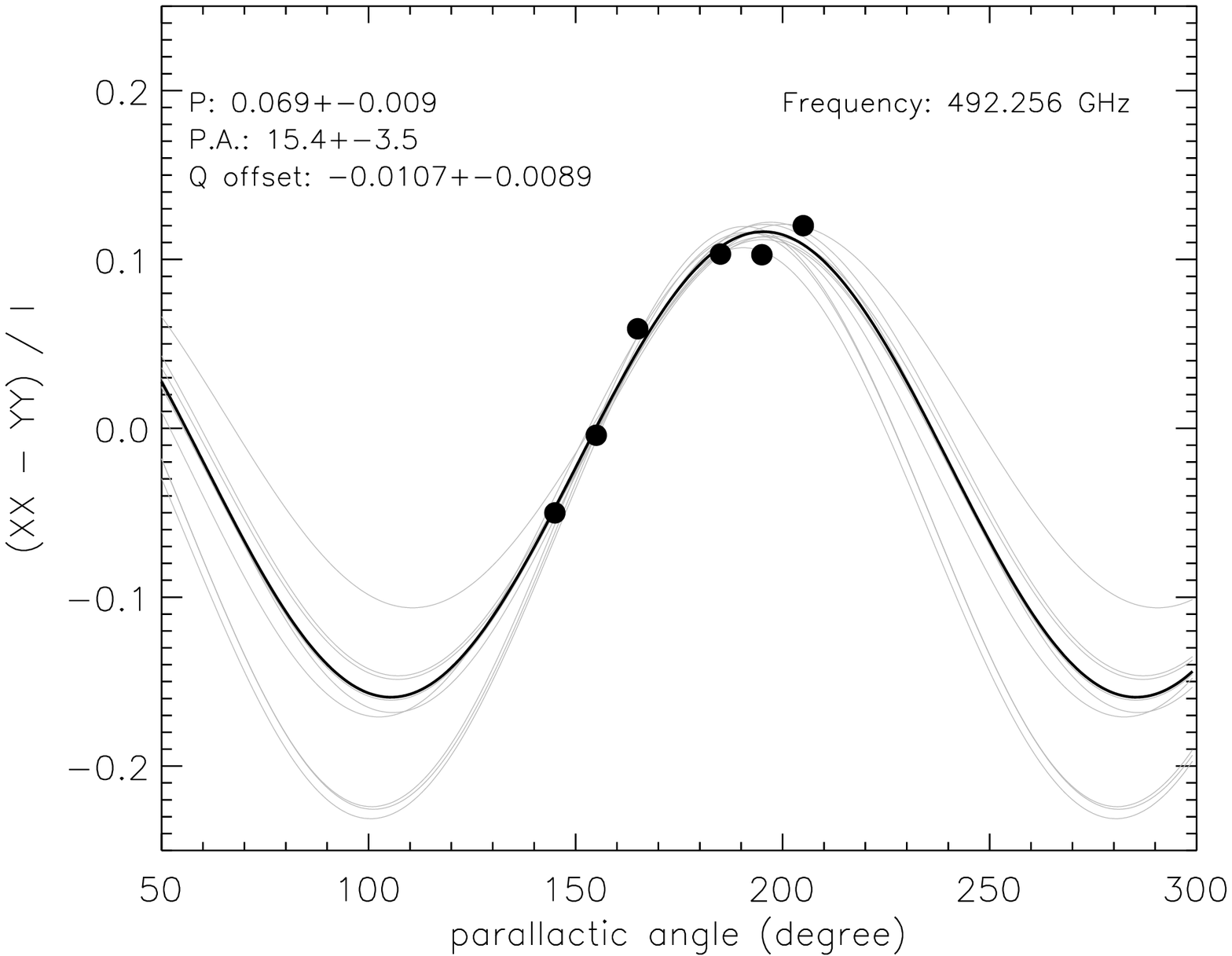} & \includegraphics[width=9.5cm]{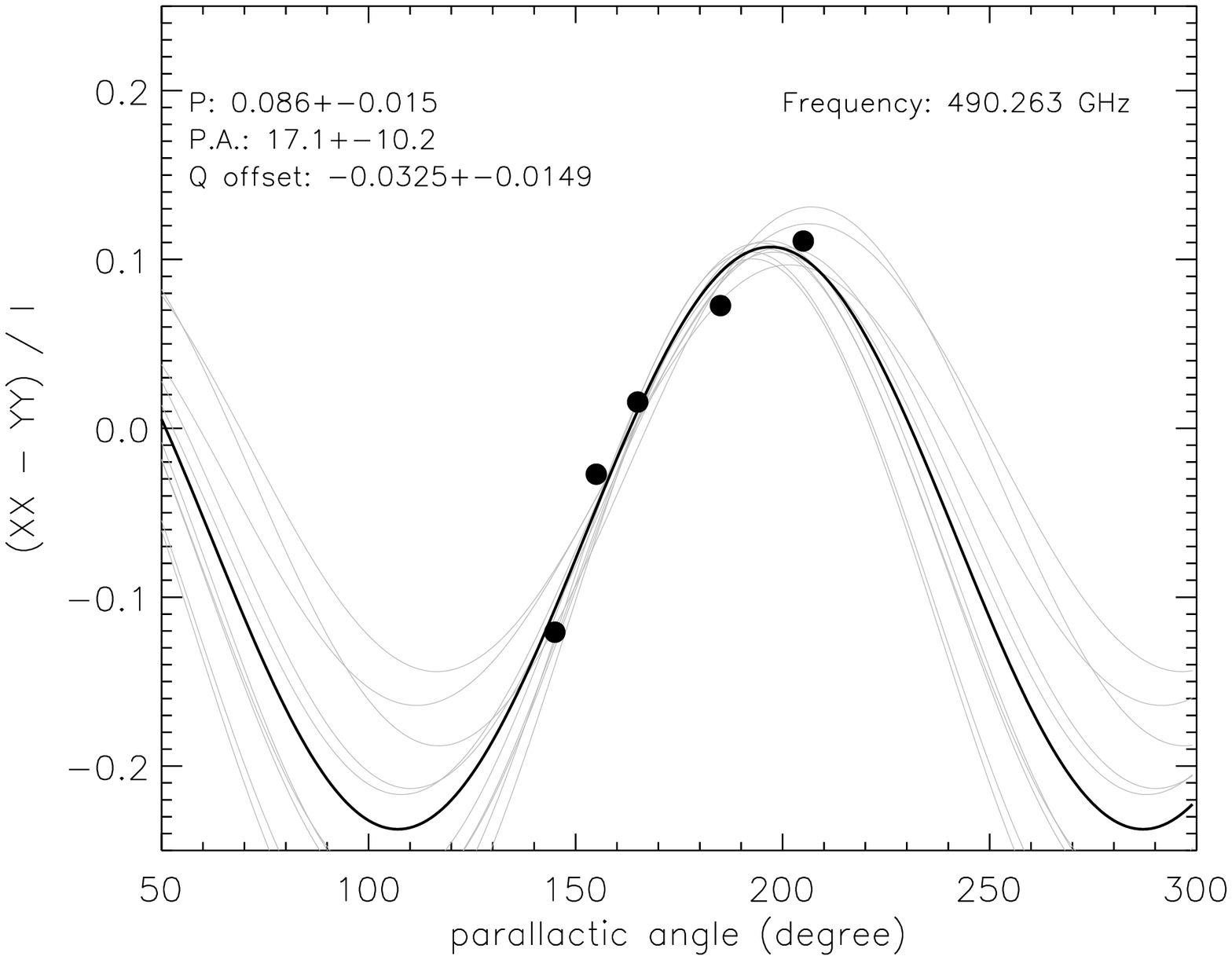} \\
\includegraphics[width=9.5cm]{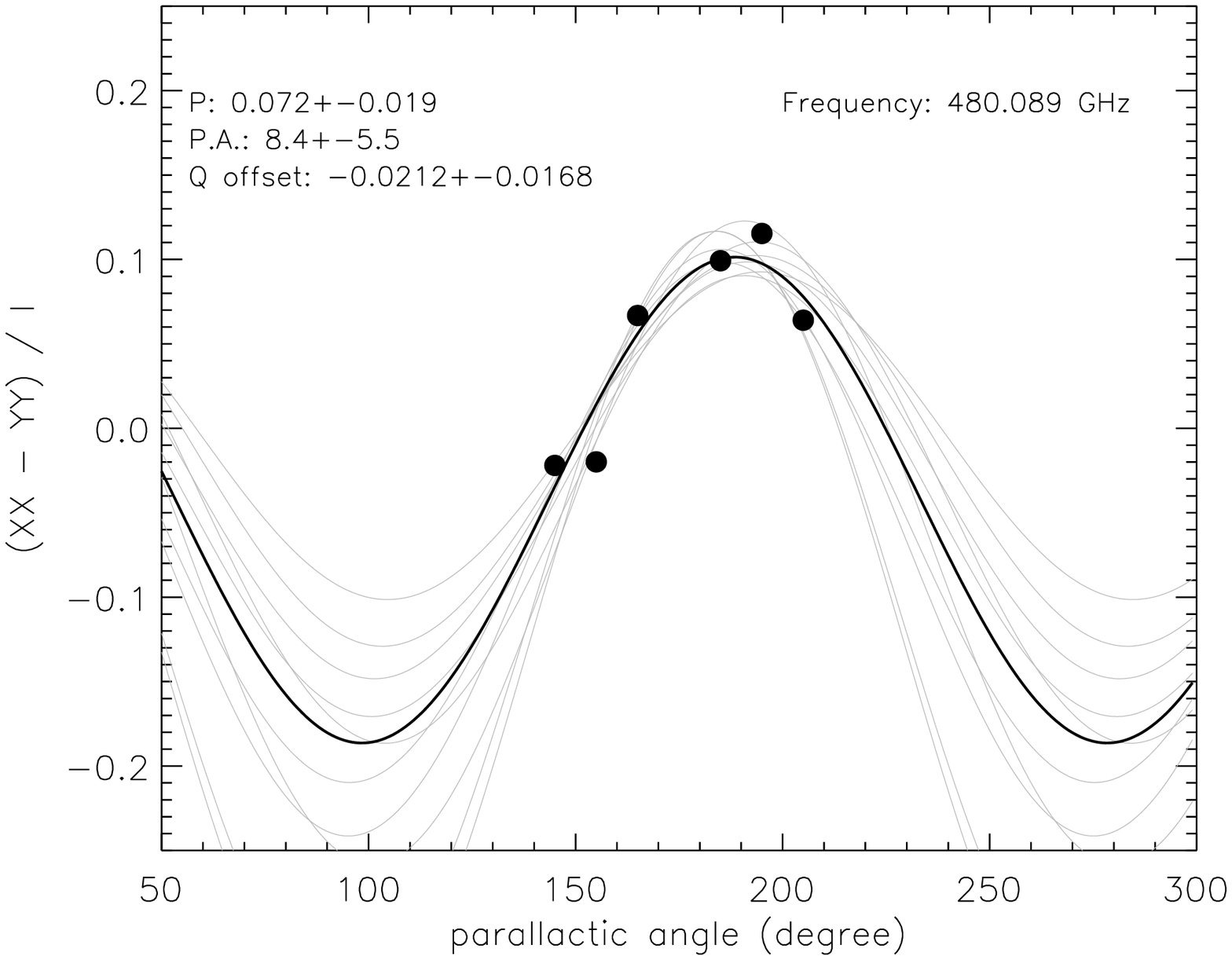} & \includegraphics[width=9.5cm]{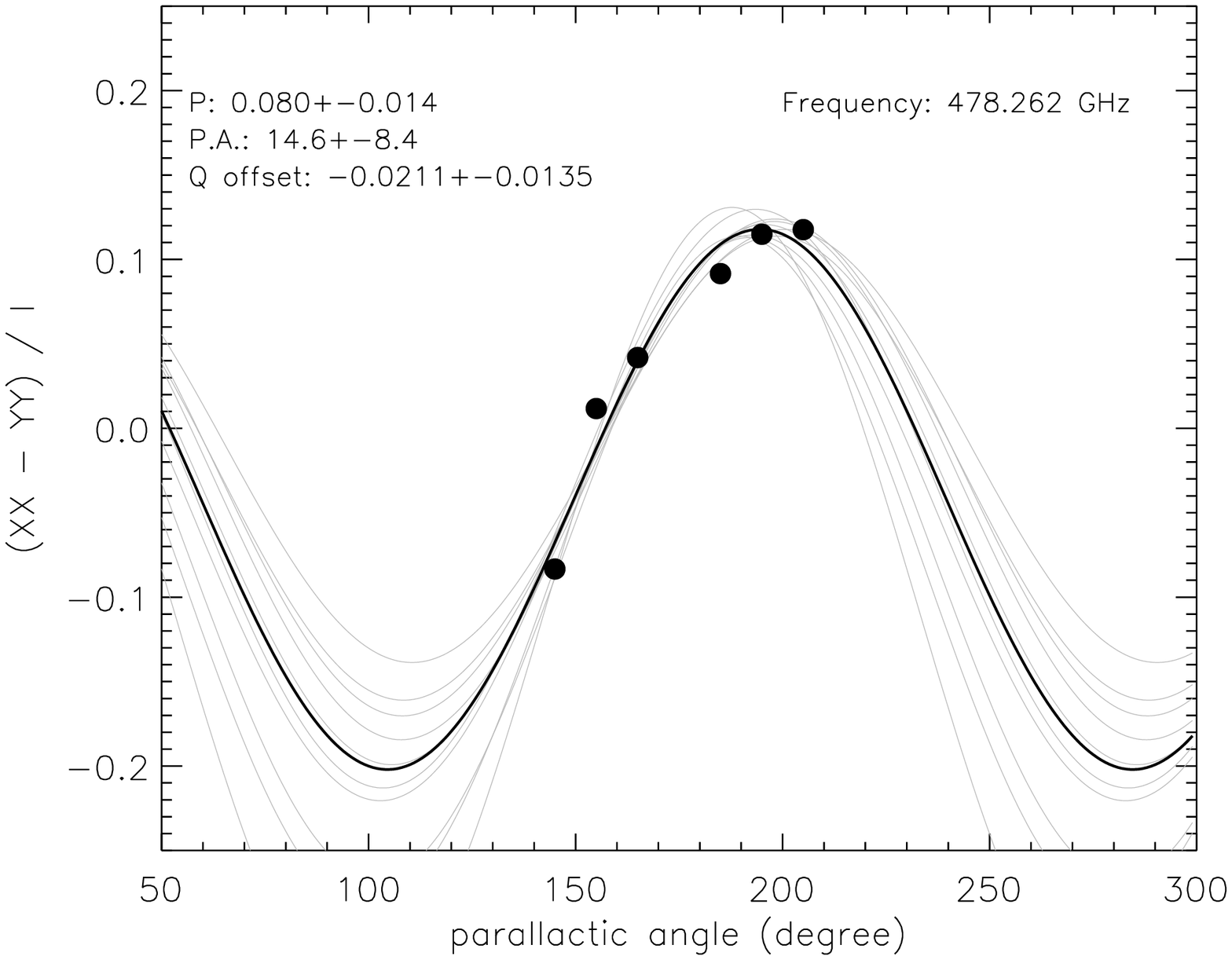} \\
\end{tabular}
\caption{Fittings of the (XX-YY)/I intensity ratio of quasar J1744-3116, to determine polarization percentages and polarization position angles. {\bf Observations from individual of the four spectral windows are presented in separated panels.} The best fits of polarization percentage, polarization position angle (in the receiver frame; P.A.), and a constant normalized Stokes Q offsets (Q offset), are provided in the upper left of each panel, which are represented by a black curve. For each observed frequency, errors of fitted quantities were determined by one standard deviation of fittings of 1000 random realizations of noisy data. Gray lines in each panel plot every 100 of the random realizations.}
\label{fig:j1744}
\end{figure*}

\begin{figure}
\hspace{-0.5cm}
\includegraphics[width=10cm]{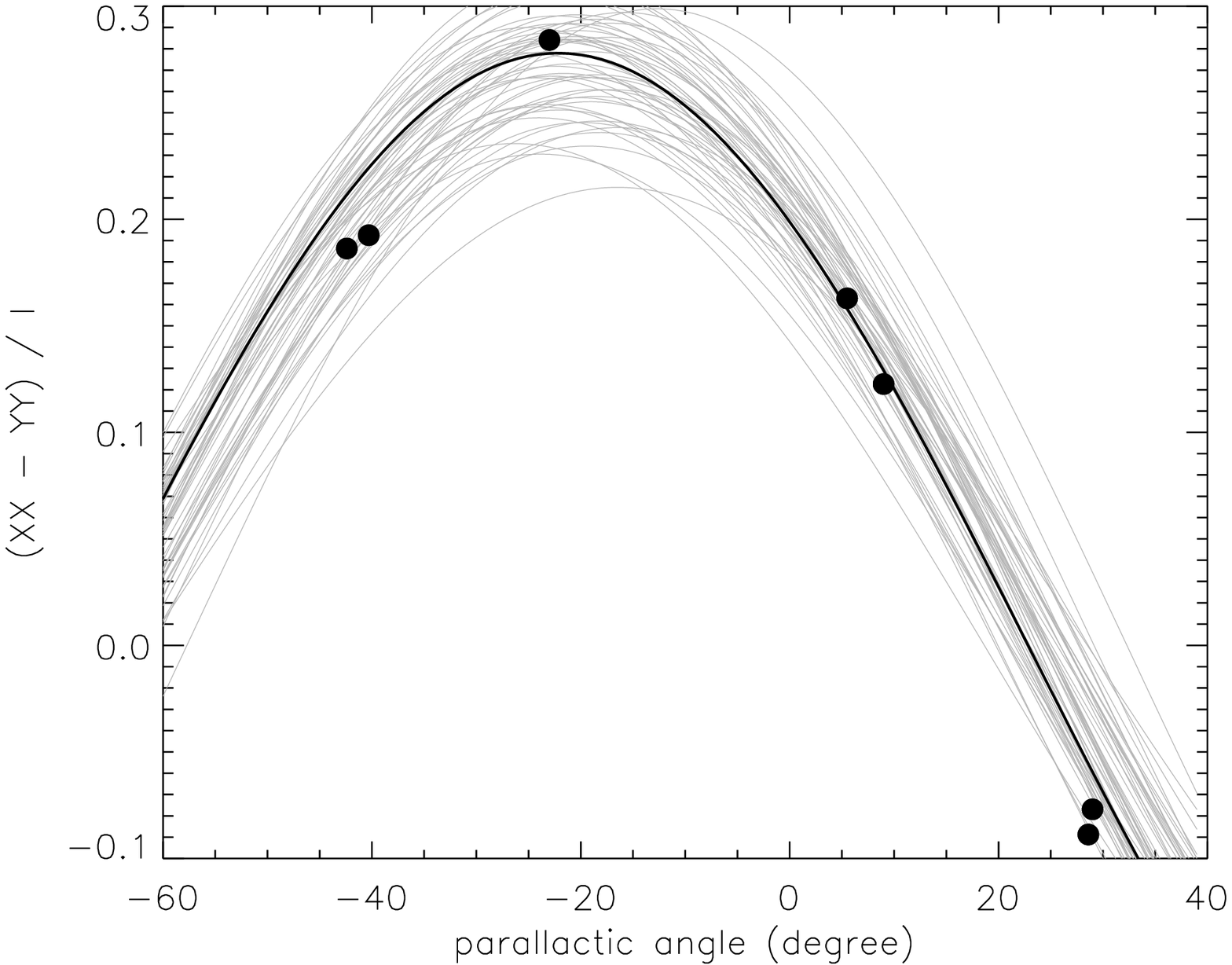}
\caption{The normalized intensity difference of the XX and YY correlations of Sgr A*, observed by the ALMA 12m-Array (symbols), and a black curve representing our best fit to these data. The constant polarization percentage and polarization position angle obtained from our best fit model are 14\%$\pm$1.2\% and 158$^{\circ}$$\pm$3$^{\circ}$, respectively. Gray curves show 50 independent random realizations of models with constant polarization percentage and polarization position angle, which characterize the error bars we give. We caution that these quantities are not fully constrained without the measurements of the XY and YX correlations.}  
\label{fig:stokesfit}
\vspace{-0.3cm}
\end{figure}

The Stokes I intensity of the Sgr A* may be varying with time, however, cannot be clearly distinguished given our present flux calibration accuracy (Figure \ref{fig:xxyytime}).
In addition, we find that Sgr A* and the gain calibrator J1744-3116 have several times higher fractional amplitude differences between the XX and the YY correlations, than that of the continuum emission of Titan.
From the $<$100 meter baselines, the XX/YY flux ratios of Titan measured from spw 0, 1, 2, and 3, are 0.99, 0.98, 1.0, and 1.0, respectively.
The relative amplitude differences of Sgr A* and J1744-3116 cannot be attributed to decoherence due to phase errors.  
The observed XX and YY amplitudes of the gain calibrator J1744-3116 can be fitted to Stokes I = 252 mJy, with 7.9\%$\pm$0.9\% polarization in position angle P.A. = 14.1$^{\circ}$$\pm$4.2$^{\circ}$, and a constant normalized Stokes Q offset $\delta$=$-$0.02$\pm$0.02, which may be caused by amplitude calibration errors or polarization leakage (Figure \ref{fig:j1744}).
However, the XX and YY amplitudes of Sgr A*, obtained from the inner 7 mosaic fields, do not vary smoothly with parallactic angle.
To first order, taking the intensity ratio of these two correlations removes the total intensity variations. 
Plotting the XX to YY intensity ratio versus parallactic angle {\bf from the 12m-Array observations} shows a peak at a parallactic angle of $-$22$^{\circ}$, with an intensity ratio close to 1 around parallactic angle +20$^{\circ}$,
From a least square fit to constant polarization position angle, the measured XX to YY intensity ratios for Sgr A* are consistent with the polarization percentage of $\sim$14\%$\pm$1.2\% and a position angle of $\sim$158$^{\circ}$$\pm$3$^{\circ}$ (Figure \ref{fig:stokesfit}).
For comparison, previously measured polarization position angles at 340 GHz were $\sim$136$^{\circ}$-163$^{\circ}$, and showed variations on daily timescales (Marrone et al. 2006a).
The imperfect fits of Figure \ref{fig:stokesfit}, if not due to calibration issues (more discussion in Section \ref{chap_discussion}), may be attributed to time variation in the polarization percentage and position angles during the period of our ALMA observations. 
However we cannot easily verify this without observing and calibrating the XY and YX cross correlations.
We refer to Bower et al. (2003) and Marrone et al. (2006a) for the observational evidence and discussion of polarization percentage variability at the 230 and the 340 GHz bands.
We refer to Eckart et al. (2006), Fish et al. (2009), Zamaninasab et al. (2010) and references therein, for modeling frameworks of the polarized emission.

\begin{figure*}
\hspace{0.15cm}
\begin{tabular}{p{8cm} p{8cm}}
\includegraphics[width=9.5cm]{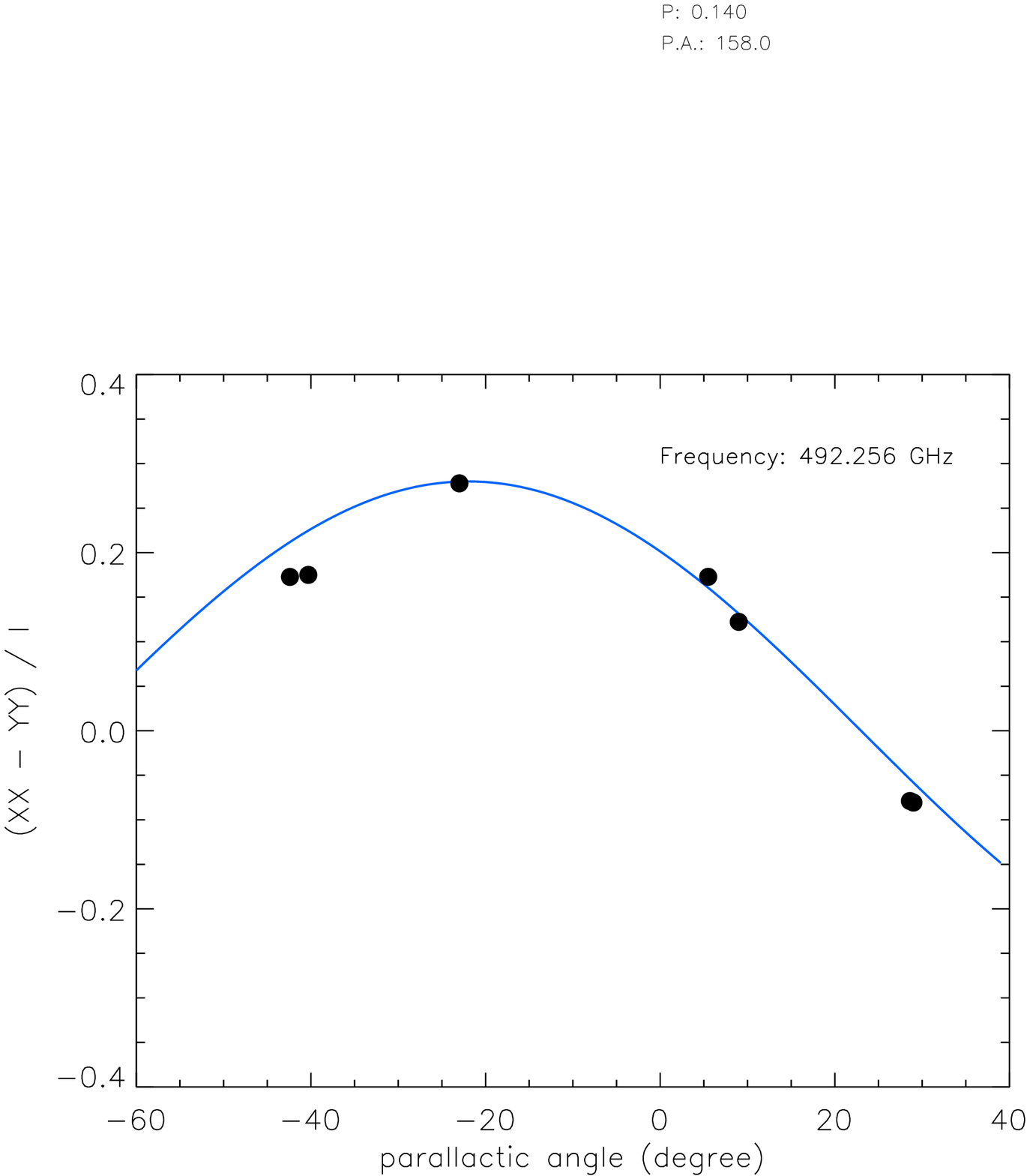} & \includegraphics[width=9.5cm]{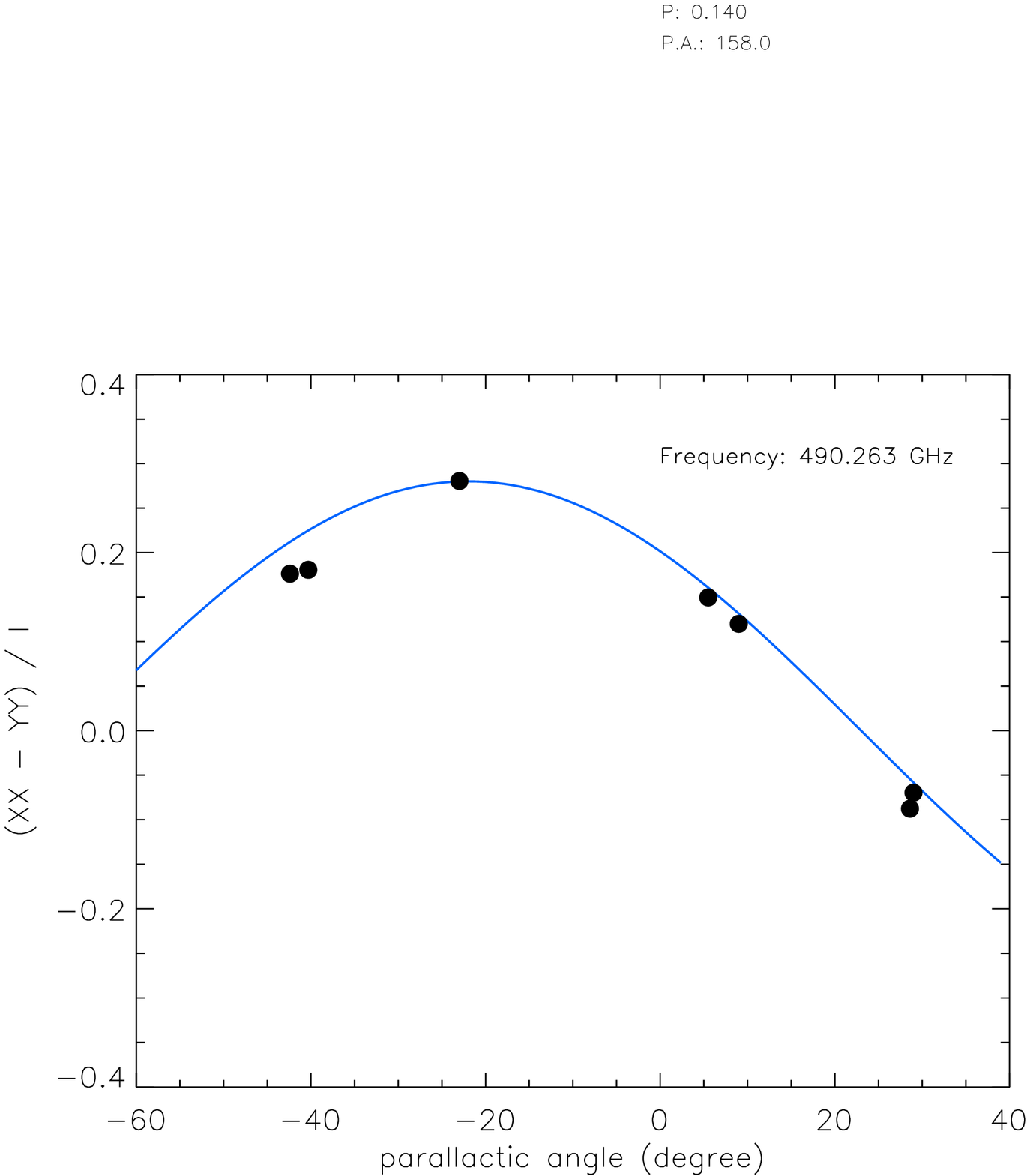} \\
\includegraphics[width=9.5cm]{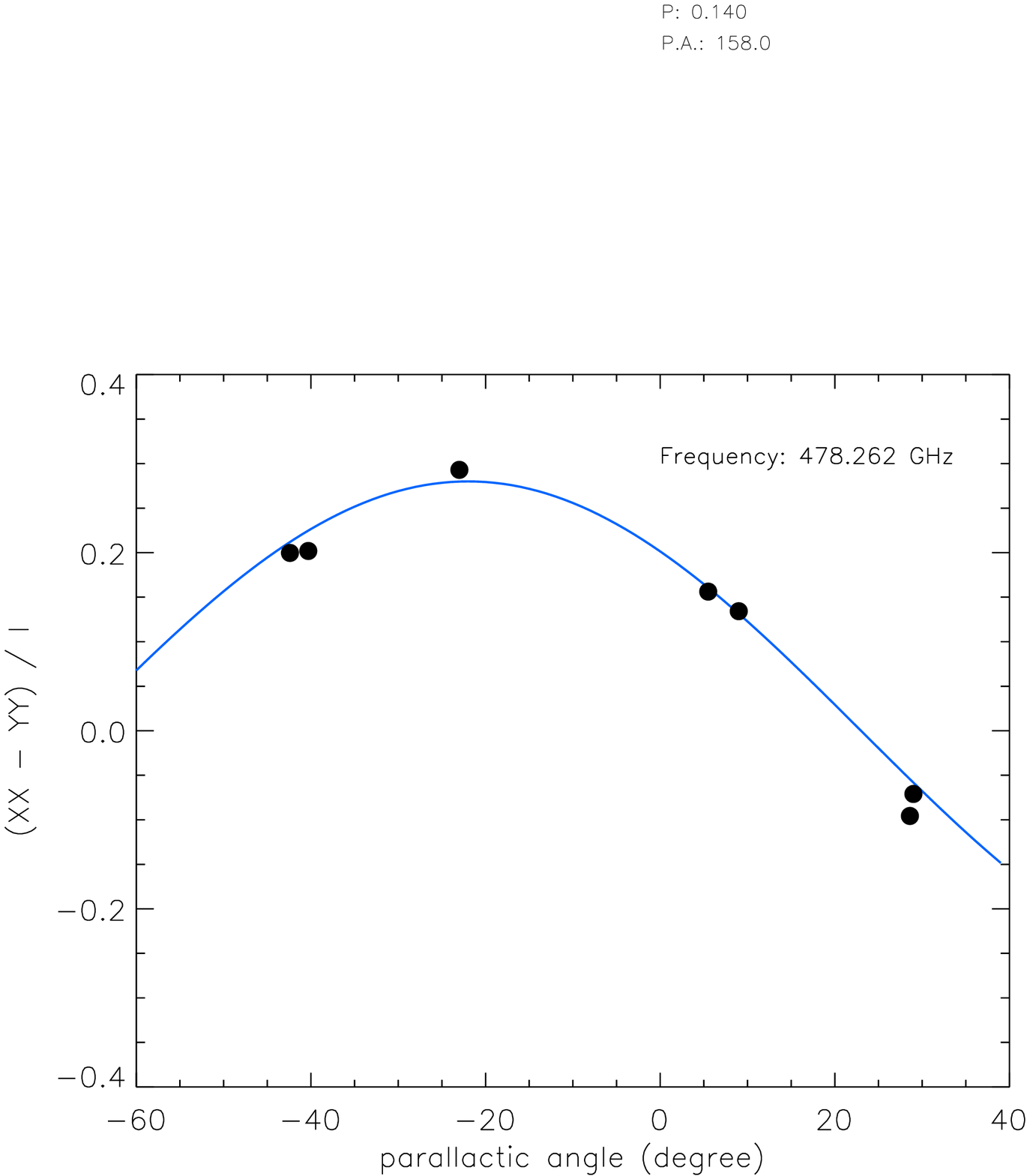} & \includegraphics[width=9.5cm]{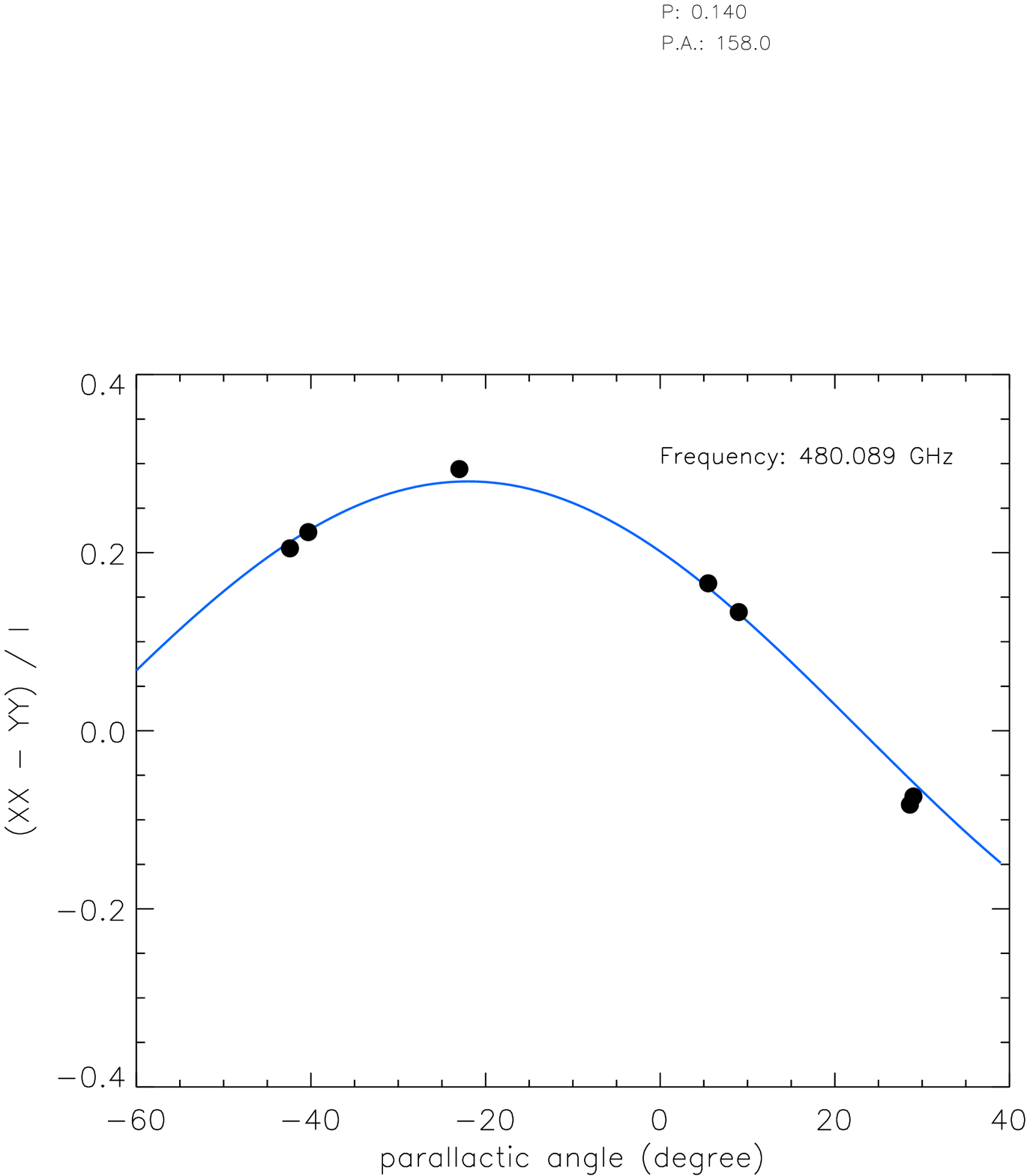} \\
\end{tabular}
\caption{ {\bf The (XX-YY)/I intensity ratio of Sgr A*. Observations from individual of the four spectral windows are presented in separated panels. Blue lines show the best fit in Figure \ref{fig:stokesfit}.} }
\label{fig:sgrAspws}
\end{figure*}

To determine whether there might be a spurious polarization signal due to the heterogeneity of dishes in the 12m-Array, we split the 12m-Array visibility data into subsets containing only correlation products between the DA antennas, only correlation products between the DV antennas, and a subset containing all correlation products between the DA and the DV antennas. 
We obtained identical measurements from these three subsets.
Therefore, we are convinced that there is no detectable spurious polarizations due to different DA and DV antenna designs. 
There were only two PM antennas in our 12m-Array observations, so we could not reliably check the cross products independently.
Nevertheless, we found that including or not including the PM antennas does not significantly change our measurements.
The XX and YY intensity differences of Sgr A* observed from the four spectral windows are also consistent {\bf(Figure \ref{fig:sgrAspws})}.

\subsection{Spectral line data} 
\label{subsec:line}
We are not aware of any mechanism which can uniformly polarize C\textsc{i} line emission to a high percentage over our mosaic field of view. 
Thermal continuum emission of Titan is also not known to be polarized.
Therefore, we use these observations to gauge the magnitude of spurious polarization caused by the offset of antenna response in XX and YY, and polarization leakage.

We used the Miriad task {\tt imdiff} to systematically estimate the multiplicative factor that minimizes the difference between the XX and YY synthesized images of C\textsc{i} in a maximum likelihood sense. 
We note that this multiplicative factor ($F_{XX}^{CI}(\nu, t)/F_{YY}^{CI}(\nu, t)$, hereafter) can depend on $v_{lsr}$ and time.
To avoid the high noise at the edge of the 12m-Array mosaic field, we limited the derivation of $F_{XX}^{CI}(\nu, t)/F_{YY}^{CI}(\nu, t)$ to a box-shaped region containing the most significant C\textsc{i} emission.
The coordinates of the bottom left and top right corners of this region are R.A. (J2000) =17$^{\mbox{\scriptsize{h}}}$45$^{\mbox{\scriptsize{m}}}$41$^{\mbox{\scriptsize{s}}}$.332, and decl. (J2000) =$-$29$^{\circ}$00$'$56$''$.77 and R.A. (J2000) =17$^{\mbox{\scriptsize{h}}}$45$^{\mbox{\scriptsize{m}}}$38$^{\mbox{\scriptsize{s}}}$.885, and decl. (J2000) =$-$28$^{\circ}$59$'$56$''$.77, respectively.
We verify that using the full images for estimating $F_{XX}^{CI}(\nu, t)/F_{YY}^{CI}(\nu, t)$ does not change the results, although it can change the noise behavior. 
We also measured the XX to YY amplitude ratio of the 12m-Array continuum observations of Titan, using the same method.
The continuum emission from Titan shows a $\sim$3\% intensity difference between the XX and the YY correlations. 
The XX to YY continuum intensity ratios of both Sgr A* and Titan are shown in Figure \ref{fig:xxyytime}.

\begin{figure}
\vspace{-1cm}
\hspace{-1cm}
\hspace{0.15cm}
\begin{tabular}{p{7.5cm} }
\includegraphics[width=9.5cm]{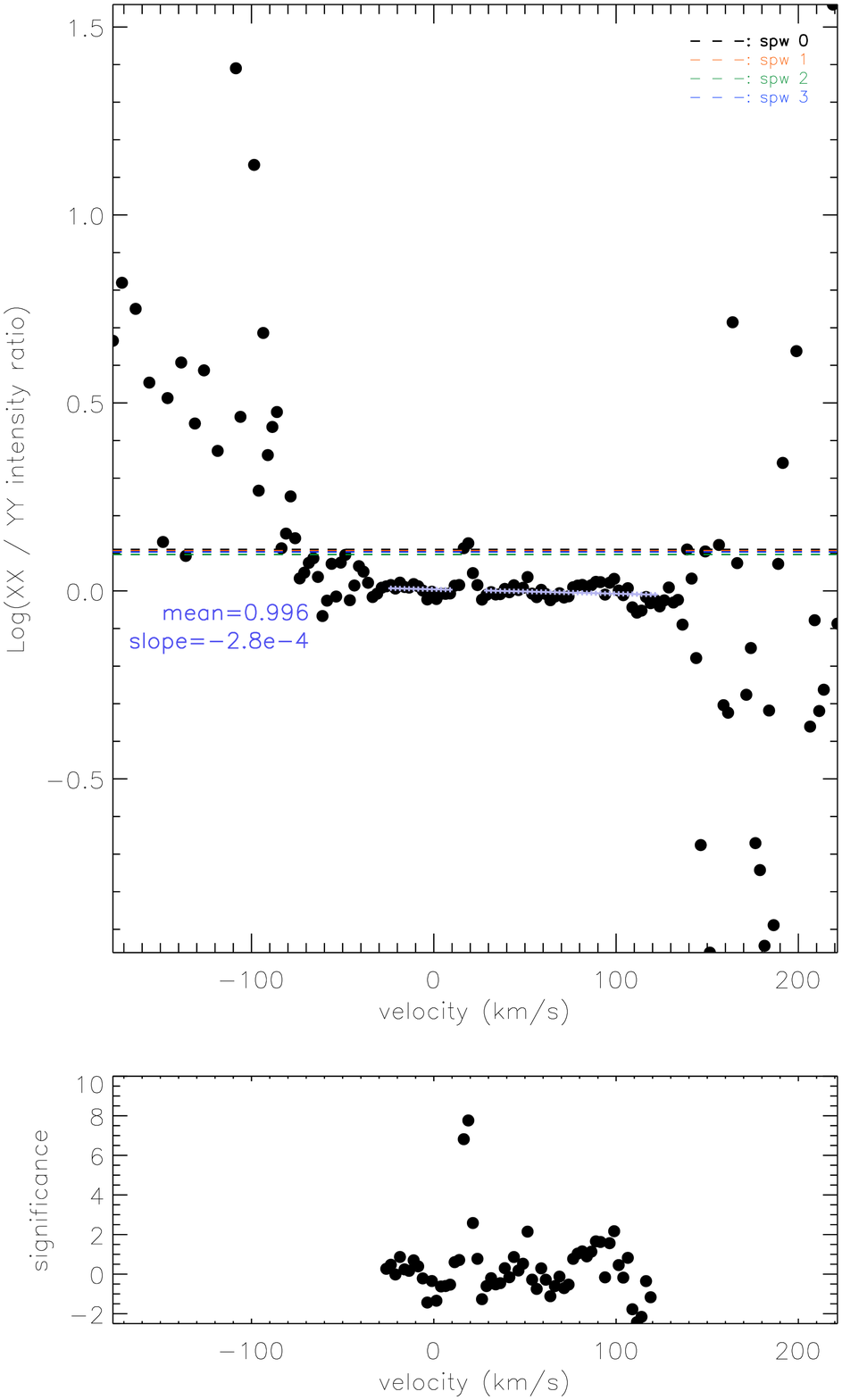} \\
\end{tabular}

\vspace{-1.0cm}
\caption{\small{Top panel shows the derived intensity ratio between the XX and the YY correlations ($F_{XX}^{CI}(\nu)/F_{YY}^{CI}(\nu)$) from C\textsc{i} line velocity channel {\it synthesized} image, as a function of $v_{lsr}$.
We overplotted the XX and YY intensity ratios derived from the continuum data of the inner 7 12m-Array mosaic fields (dashed lines, color coding the four spectral windows). 
The large scattering in the redshifted and blueshifted ends are because we did not detect C\textsc{i} emission or absorption and therefore fittings of $F_{XX}^{CI}(\nu)/F_{YY}^{CI}(\nu)$ did not converge. 
We performed linear regression for the high S/N spectral channels, and derived the $\pm$1 standard deviation ($\sigma_{\nu}^{CI}$) of the differences from the regression line. 
Results of linear regression is shown by light blue symbols.
We plot  ($F_{XX}^{CI}(\nu)/F_{YY}^{CI}(\nu)$) / ($\sigma_{\nu}^{CI}$) in the bottom panel.
}}
\label{fig:xxyyfreq}
\vspace{-0.1cm}
\end{figure}

We define $\int F_{XX}^{CI}(\nu, t) d\nu$ / $\int d\nu$ = $F_{XX}^{CI}(t)$, and $\int F_{XX}^{CI}(\nu, t) dt$ / $\int dt$ = $F_{XX}^{CI}(\nu)$.
In practice, we measured $F_{XX}^{CI}(t)$ / $F_{YY}^{CI}(t)$ of the 12m-Array observations from spectral channels which are dominated by C\textsc{i} {\it emission} (the case in which it is dominated by absorption is described further below), for each of the target source scans (i.e. every time period bracketed by two adjacent gain calibration scans).
$F_{XX}^{CI}(t)$ / $F_{YY}^{CI}(t)$ for the ACA observations were measured in the same way, but over the entire ACA observing period.
We also measured $F_{XX}^{CI}(\nu)$ / $F_{YY}^{CI}(\nu)$ for every 2.5 km\,s$^{-1}$ wide velocity channels, by averaging over all 12m-Array integrations. 
However, we were not able to obtain a meaningful constraint of $F_{XX}^{CI}(\nu)$ / $F_{YY}^{CI}(\nu)$ from the ACA observations, due to their limited sensitivity.
Figure \ref{fig:xxyyfreq} shows the measured $F_{XX}^{CI}(t)$ / $F_{YY}^{CI}(t)$ and $F_{XX}^{CI}(\nu)$ / $F_{YY}^{CI}(\nu)$ from our observations.

\begin{figure*}[h!]
\hspace{-0.75cm}
\hspace{0.15cm}
\includegraphics[width=18cm]{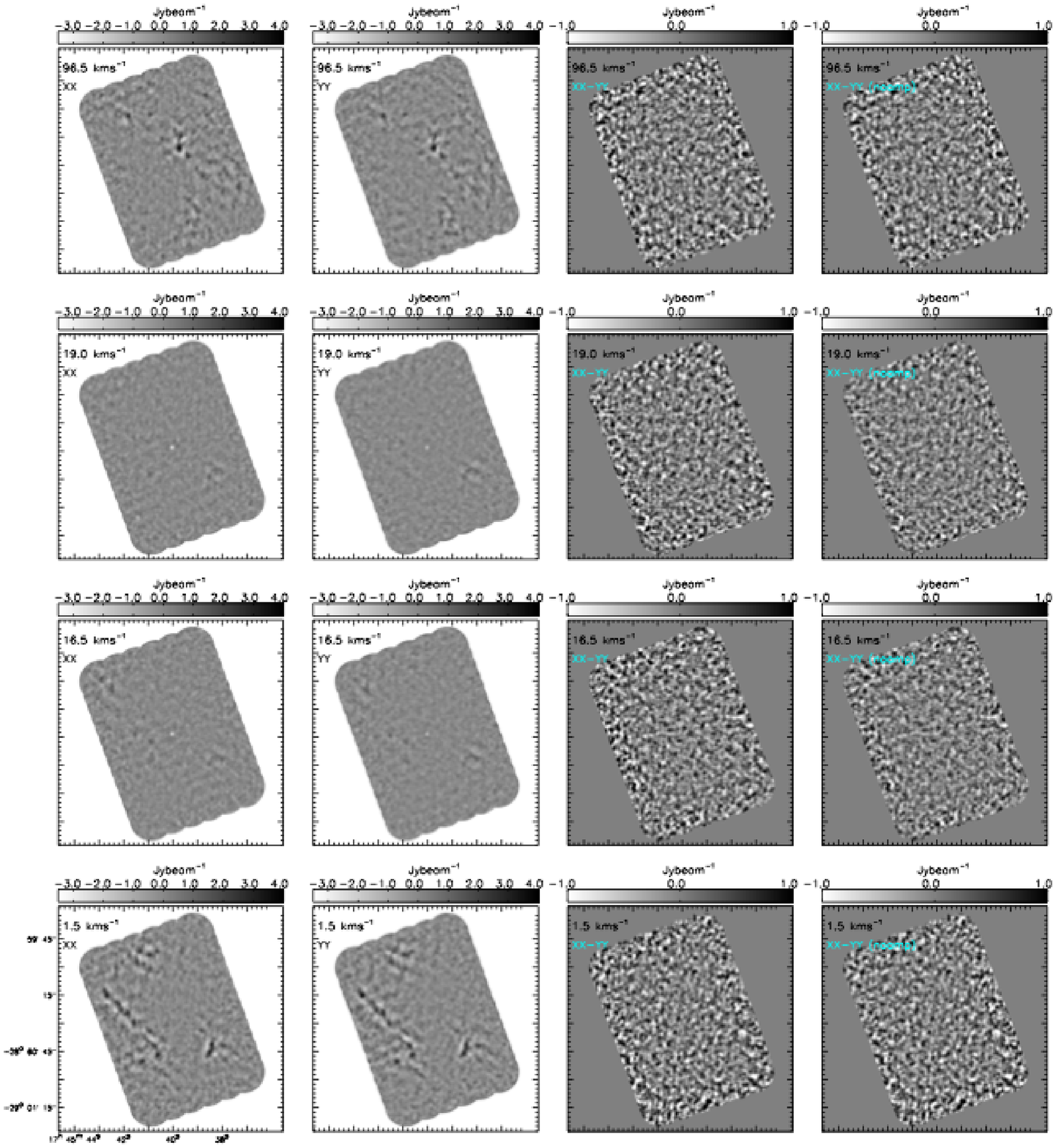}




\caption{\small{Left to right columns show the XX, YY , XX - YY, and XX - $a^{\mbox{\scriptsize{CI}}}(\nu)$YY {\it synthesized} intensity maps of selected velocity channels. 
These images were generated using the 12m-Array data only (Section \ref{chap_obs}), and were further tapered by a Gaussian weighting function of FWHM = $1\farcs5$.
The C\textsc{i} line emission close to $v_{lsr}$$\sim$20 km\,s$^{-1}$ is subject to foreground absorption, such that the images are dominated by the point-like absorption signature against Sgr A*. Extended C\textsc{i} emission features surrounding Sgr A* are present away from $v_{lsr}$$\sim$20 km\,s$^{-1}$. 
}}
\label{fig:images}
\vspace{0.5cm}
\end{figure*}

Extended emission from the C\textsc{i} line is detected in channels over a range of velocities, following a similar velocity field to that of the molecular circumnuclear disk (Guesten et al. 1987; Wright et al. 2001; Liu et al. 2012, 2013, and references therein). Examples of the C\textsc{i} line velocity channel {\it synthesized} images from the 12m-Array observations, are given in Figure \ref{fig:images}. The CI emission will be discussed in more detail in a separate paper (Liu et al. in prep.)
However, we found that for several velocity channels around $v_{lsr}$$\sim$20 km\,s$^{-1}$, the extended CI line emission from the Galactic center is nearly completely absorbed by foreground gas. 
In these channels, the dominant feature is absorption against the continuum emission of Sgr A*, which is not spatially resolved by our observations.

\begin{figure}
\vspace{-0.3cm}
\includegraphics[width=9.5cm]{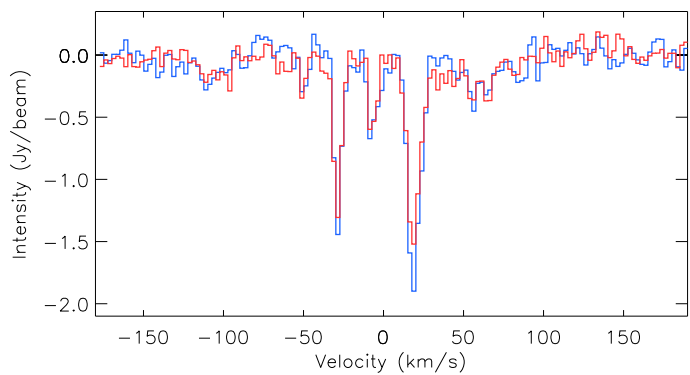}
\caption{The C\textsc{i} line spectra taken from a single synthesized beam area surrounding the Sgr A*. Blue and red lines present the XX and YY polarizations.}
\vspace{-0.3cm}
\label{fig:spectrum}
\end{figure}

At the same velocity as the absorption we detect a local maximum of $F_{XX}^{CI}(\nu)$ / $F_{YY}^{CI}(\nu)$ (Figure \ref{fig:xxyyfreq}).
The local peak value of $F_{XX}^{CI}(\nu)$ / $F_{YY}^{CI}(\nu)$ is $\sim$1.3 (or 0.11 in logarithm).
This peak value of $F_{XX}^{CI}(\nu)$ / $F_{YY}^{CI}(\nu)$ is consistent within 1$\sigma$ with the XX and YY continuum intensity ratio of Sgr A*, measured from the inner 7 fields mosaic of the  12m-Array observations.
In fact, the three most prominent absorption line features of C\textsc{i} against the continuum emission of the Sgr A*, consistently present a deeper absorption in XX correlation than in YY (Figure \ref{fig:spectrum}).
In the ACA observations, the difference of the absorption line intensities between the XX and the YY correlations, are lower than the 1$\sigma$ noise level of the ACA observations. 

$F_{XX}^{CI}(\nu)$ / $F_{YY}^{CI}(\nu)$ is close to 1 in the remaining velocity channels with significant emission.
The standard deviation of $F_{XX}^{CI}(\nu)$ / $F_{YY}^{CI}(\nu)$, $\sigma_{\nu}^{CI}$, measured from velocity channels away from $v_{lsr}$ = 20 km\,s$^{-1}$ (Figure \ref{fig:xxyyfreq}), is 0.043.
For the velocity range in which we significantly detected C\textsc{i}, the value of [Max($F_{XX}^{CI}(\nu)$ / $F_{YY}^{CI}(\nu)$ )  -  Mean($F_{XX}^{CI}(\nu)$ / $F_{YY}^{CI}(\nu)$) ]  / $\sigma_{\nu}^{CI}$ is 7.8 (Figure \ref{fig:xxyyfreq}). 
We have visually inspected the XX and YY intensity maps ($I^{XX}(\nu, t)$, $I^{YY}(\nu, t)$)\footnote{Here $I^{XX, YY}(\nu, t)$ refers to the intensity maps taken at a specific time $t$, rather than time variation of intensity at any specific position.}, and the residual $R(\nu, t)$ $\equiv$ $I^{XX}(\nu, t)$ - ($F_{XX}^{CI}(\nu, t)$ / $F_{YY}^{CI}(\nu, t)$)$\times$$I^{YY}(\nu, t)$.
Based on the statistics of pixel values and the visual inspection of images, we found that $R(\nu, t)$, and its time integration, are consistent with thermal noise.
On the other hand, we found that for spectral channels away from $v_{lsr}$$\sim$20 km\,s$^{-1}$, $I^{XX}(\nu, t)$ $-$1.3$\times$$I^{YY}(\nu, t)$ presents significant (i.e. $>$3$\sigma$) features of over subtraction.

Figure \ref{fig:xxyyfreq} and \ref{fig:spectrum} may be understood considering the radiative transfer equation
$
T_{b} = (T_{ex} - T_{bg})( 1 - e^{-\tau}),
$
where $T_{b}$ is the observed C\textsc{i} brightness temperature, $T_{ex}$ is the gas excitation temperature, $T_{bg}$ is the background brightness temperature, and $\tau$ is the optical depth of gas. 
For the foreground C\textsc{i} absorption against the continuum emission of Sgr A*, it is safe to assume that $T_{ex}$ is negligible, and the gas optical depth $\tau$ is identical for the orthogonal linear polarizations X and Y. 
The assumption of the identical gas optical depth $\tau$ for the X and Y polarizations can be supported by the observed XX/YY$\sim$1 from emission line (Figure \ref{fig:xxyyfreq}).
Therefore, the C\textsc{i} absorption line ratio of the XX and YY correlations, is expected to be nearly identical to the XX/YY flux ratio of the continuum emission of Sgr A*.

\section{Discussion}
\label{chap_discussion}
The significant difference between the XX and YY correlations can be used to make a reliable determination of Stokes Q at 492 GHz.
However, lacking the cross-correlations XY and YX which were not sampled in these observations, we are not able to determine Stokes U.
Nevertheless, the ALMA observations give a meaningful lower limit on the linear polarization continuum emission from Sgr A* at this highest frequency that has been studied in polarization there with any submillimeter bands available on interferometer arrays to date.
The maximum of the intensity differences between the XX and the YY correlations observed from the $\sim$492 GHz continuum emission of Sgr A* (diff($I_{XX,YY}^{cont.}$), hereafter), implies a $\sim$14\%$\pm$1.2\% lower limit on its polarization percentage.

Potential causes of the observed diff($I_{XX,YY}^{cont.}$) are synchrotron emission from ionized gas close to Sgr A* (Falcke et al. 1998; Aitken et al. 2000; Bower et al. 2001, 2003; Marrone et al. 2006a; Bromley et al. 2001; Liu et al. 2007; Huang et al. 2009), or instrumental effects including (1) beam squint, (2) relative drifts of instrumental gain amplitude between the XX and YY correlations, (3) phase decoherence for a certain polarization, and (4) primary beam polarization.
As addressed in Section \ref{chap_obs}, we find no evidence that the decoherence due to phase errors can lead to the differences of intensities measured by the XX and YY correlations.
In addition, our analysis of the C\textsc{i} line emission has ruled out the possibilities that the relative drifts of instrumental gain amplitude as well as the effects of phase decoherence can lead to the observed diff($I_{XX,YY}^{cont.}$) in continuum emission (Section \ref{subsec:line}).
Beam squint does not apply to the observations on the central field (field 0, see Figure \ref{fig:fields}).
The observed diff($I_{XX,YY}^{cont.}$) from the other mosaic fields  also appears too big to be explained by beam squint, unless the actual primary beam response functions of the ALMA antennas seriously deviate from the present understanding.
Nevertheless, the comparisons of the diff($I_{XX,YY}^{cont.}$) taken from the pairs of fields (18, 25), (94, 101), and (133, 134), which were observed closely in time, empirically provide a limit on the scale of beam squint effects (Figure \ref{fig:xxyytime}).
On the other hand, each two exposures of these three pairs of fields show rather consistent diff($I_{XX,YY}^{cont.}$), which may indicate that there is no significant variation of polarization on the very short timescales probed by their time separations.
Primary beam polarization cannot explain the observed highest diff($I_{XX,YY}^{cont.}$) from the central field (i.e. Field 0, see Figure \ref{fig:xxyytime}), and cannot explain the frequency dependence of $F_{XX}^{CI}(\nu)$ / $F_{YY}^{CI}(\nu)$ (Figure \ref{fig:xxyyfreq}).
We are not aware of other instrumental defects which can cause similar effects, and consider polarized synchrotron emission as the most probable explanation for the diff($I_{XX,YY}^{cont.})$ we measured from Sgr A*.

\begin{figure}[h!]
\vspace{-1cm}
\hspace{-0.5cm}
\begin{tabular}{p{7.5cm} }
\includegraphics[width=10cm]{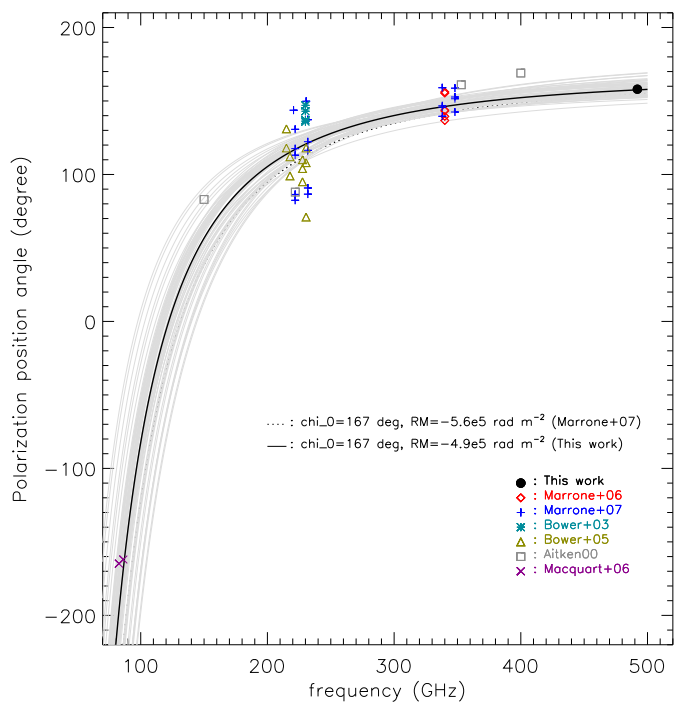} \\
\end{tabular}
\vspace{-1cm}
\caption{\small{The observed polarization position angle of Sgr A* at 492 GHz is compared with prior data from Aitken et al. (2000), Bower et al. (2003, 2005), the mean of Macquart et al. (2006), and Marrone et al. (2006a, 2007). The polarization position angles of the Macquart et al. (2006) data were unwrapped by $-$180$^{\circ}$. We overplot the mean fitted intrinsic polarization position angle the and rotation measure by Marrone et al. (2007), and the updated fit including our measurement at 492 GHz. Gray curves show 50 independent random realizations which characterize our fitting errors. We note that our fitting error is not well defined because our observations did not include the XY and YX cross correlations. We omit error bars due to the crowded data points. However, we note that the scattering of data points due to actually observed time variations (e.g. Bower et al. 2005, Marrone et al. 2006a, 2007), is more significant than the measurement errors. Therefore, the data we present in this figure are adequate for providing a sense of uncertainties in rotation measure. The same arguments are also applied to Figure \ref{fig:fracplot}.
}}
\label{fig:chiplot}
\vspace{-0.25cm}
\end{figure}

\begin{figure}[h!]
\vspace{-0.5cm}
\hspace{-0.75cm}
\begin{tabular}{p{7.5cm} }
\includegraphics[width=10cm]{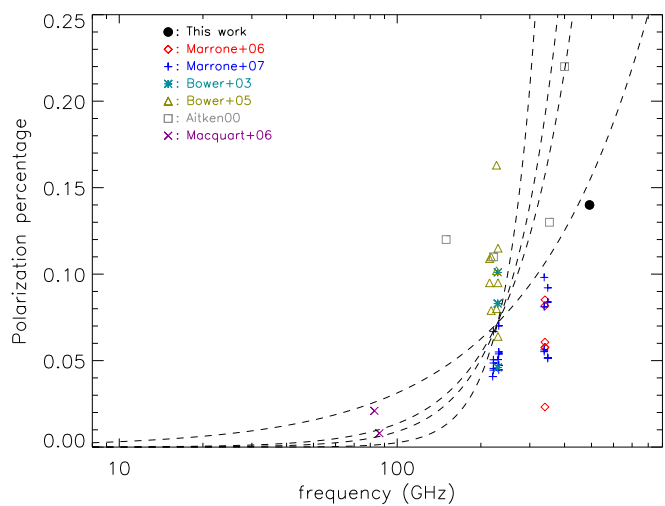} \\
\end{tabular}
\vspace{-0.3cm}
\caption{\small{The observed polarization percentage of Sgr A* at 492 GHz is plotted with prior data from Aitken et al. (2000), Bower et al. (2003, 2005), the mean of Macquart et al. (2006), and Marrone et al. (2006a, 2007). We overplot the power-law models with indices of 1.0, 2.0, 2.5 and 4, which were presented by Bower et al. (2003) with dashed lines.
}}
\label{fig:fracplot}
\end{figure}

The $\sim$14\%$\pm$1.2\% polarization percentage and 158$^{\circ}$$\pm$3$^{\circ}$ polarization position angle of the continuum emission of Sgr A* appear realistic when compared with previous (sub)millimeter observations at other frequency bands (Aitken 2000, Bower et al. 2003, 2005, Marrone et al. 2006a, 2007), despite the large time separations of these observations (Figure \ref{fig:chiplot}, \ref{fig:fracplot}).
In particular, Marrone et al. (2007) reported the fitted intrinsic polarization position angle $\chi_{0}$=167$^{\circ}$$\pm$7$^{\circ}$ and the rotation measure RM=($-$5.6$\pm$0.7)$\times $10$^{5}$ rad\,m$^{-2}$, which inplies a 155$^{+9}_{-8}$$^{\circ}$ polarization position angle at 492 GHz.
This is consistent with our new measurement of polarization position angle within 1$\sigma$.
Least square fitting of our measured polarization position angle at 492 GHz, together with the records provided by Bower et al. (2003, 2005), Macquart et al. (2006), and Marrone (2006a, 2007), yield $\chi_{0}$=167$^{\circ}$$\pm$7$^{\circ}$, and RM of ($-$4.9$\pm$1.2)$\times $10$^{5}$ rad\,m$^{-2}$, which essentially cannot be distinguished from the aforementioned fitting results of Marrone et al. (2007), and the results of $\chi_{0}$=168$^{\circ}$$\pm$8$^{\circ}$ and RM=($-$4.4$\pm$0.3)$\times $10$^{5}$ rad\,m$^{-2}$ given by Macquart et al. (2006).
We note that there is a discrepancy between the intrinsic polarization position angle determined with millimeter and submillimeter band observations, and that determined with near infrared observations (Eckart et al. 2006; Shahzamanian et 
al. 2015). 
Assuming a thin Keplerian rotating disk geometry of the accretion flow, and the toroidal magnetic field perpendicular to the rotating axis, this nearly 90$^{\circ}$ flip of polarization position angle may be interpreted by the spatially (projected) shifted dominant polarization emission area, when the observations move gradually from the optically thicker (low frequency) to the optically thinner (high frequency) regime (e.g. Bromley et al. 2001; Liu et al. 2007; Huang et al. 2009).
Therefore, at which exact frequency the 90$^{\circ}$ polarization position angle flip occurs, will provide a particular important constraint on the property of the accretion flow model.
By comparing the Stokes I flux we detected at 492 GHz with the previous observations at lower frequencies (Marrone et al. 2006b), we found that the 492 GHz emission is very likely to be in the transition from the optically thick to the optically thin regime of the spectrum.
Our 492 GHz measurement does not yet present the suggested 90$^{\circ}$ flip of polarization position angle, which may suggest that the blueshifted side of the accretion flow does not yet fully dominate the polarized emission at this observing frequency (Huang et al. 2009).
Nevertheless, our observing frequency may not be high enough to research the turning point of polarization position angle, which is expected to be $>$1 THz in some recent radiative transfer modelings (Liu et al. 2007; Huang et al. 2009).
In addition, the comparison of the Stokes I fluxes is subject to the large time separations of those measurements.
Therefore, whether our 492 GHz observations were indeed probing the optically thin regime is uncertain.
Resolving the nature of this discrepancy will require future coordinated monitoring observations.

We point out that the polarization position angle observed in the 230 GHz band is reported to present a larger time variability than that observed in the 340 GHz band (see also Figure \ref{fig:chiplot}).
Bower et al. (2005) favored an interpretation in which the variation is attributed to variations in the medium through which the polarization propagates (i.e. the variation of rotation measure), and thereby proposed a scenario of a hot and turbulent accretion flow. 
On the contrary, Marrone et al. (2007) argued that the observed time variation of the polarization position angle is more likely due to the variation of the emission source. 
Since we only have a single epoch of observations at 492 GHz, it is probable that the consistency of our observed polarization position angle with the extrapolation of the previous observations is merely a coincidence. 
We cannot yet distinguish between these two proposed scenarios, which require future multi-epoch observations.
We note, however, that these two scenarios are not mutually exclusive.

As indicated in recent studies, Sgr A* is believed to be the source for the major events episodically along with large flares emitting luminosity up to 10$^{41-42}$ erg\,s$^{-1}$. 
The time-interval between these events is about 100 year suggested by the front of fluorescent X-ray propagating away from Sgr A* (Ponti et al. 2010; Clavel et al. 2013). 
Such extraordinary X-ray flares are also expected from the statistical analysis of the flux-density fluctuations observed in the past decades in the near IR band (Witzel et al 2012). 
In comparison to the measurements made about 10 years ago, our new measurements of the rotation measure with the ALMA may imply that both the accretion rate to and the magnetic configuration around Sgr A* have not been significantly changed in the past decade. 
No extraordinary flares have been found from the monitoring programs in multiple wavelengths from radio, submillimeter, IR and X-ray launched in the past decade. 
Our current results may be expected given this inactivity.
We note that the total flux may have varied by $\sim$50\% during our observations (see Figure \ref{fig:xxyytime}).
This might also result from pointing errors or other calibration problems in these observations.
More frequent calibrations in future observation will  be more robust for addressing this point.

Finally, our simple analysis technique may also be applied to ALMA observations of quasars used for gain calibrations, in order to generate a large database of rotation measures.
A similar technique has been used to estimate rotation measure of PKS 1830-211.
For details see Marti-Vidal et al. (2016).

\section{Conclusions}\label{chap_conclusion}

We have performed Band 8 (479-482 GHz; 489-493 GHz) mosaic observations towards the Galactic center, using the ALMA 12m-Array and the ACA.
The observed Stokes I flux of Sgr A* at 492 GHz is 3.6$\pm$0.72 Jy.
We {\bf hypothesize} that the continuum emission of Sgr A*, and the C\textsc{i} absorption line against Sgr A*, exhibit substantial intensity differences between the XX and the YY correlations.
However, the XX and YY intensities of the C\textsc{i} line emission are essentially identical, at all velocity channels for which there is significant emission and over the entire time period of the 12m-Array observations.
The maximum value of the observed intensity differences from Sgr A* implies a $\sim$14\%$\pm$1.2\% lower limit on the polarization percentage.
A comparable or higher polarization percentage of the continuum emission of Sgr A* is expected from prior observations at other frequencies (Bower et al. 2003, 2005).
The intrinsic polarization position angle we derived from the observed XX to YY intensity ratios is $\sim$167$^{\circ}$, which is surprisingly, in good agreement with the polarization position angles reported by the SMA observations at 230-340 GHz about one decade ago (Marrone et al. 2006a, 2007).
Therefore, we attribute the observed intensity differences to linearly polarized synchrotron emission from hot ionized gas immediately surrounding Sgr A*.
We found that the polarization percentage at our observing frequency may be varying over the time period of our 12m-Array observations.
Improved constraints on polarization will require new measurements that include the XY and YX correlations. 
We also detected 7.9\%$\pm$0.9\% polarization in position angle P.A. = 14.1$^{\circ}$$\pm$4.2$^{\circ}$ from the gain calibration quasar J1744-3116, which was observed at the same night with Sgr A*.


\begin{acknowledgements}
We thank our referee for the very precise and useful opinions.
HBL thanks ASIAA for support.
HBL thanks Yu-Nung Su for the help when organizing the observational proposal; and thanks Lei Huang for some basic discussion made in 2004-2006.
This paper makes use of the following ALMA data: ADS/JAO.ALMA 2013.1.00071.S. 
ALMA is a partnership of ESO (representing its member states), NSF (USA) and NINS (Japan), together with NRC (Canada) and NSC and ASIAA (Taiwan), in cooperation with the Republic of Chile. 
The Joint ALMA Observatory is operated by ESO, AUI/NRAO and NAOJ.
We thank Aaron Evans and Todd Hunter for providing information about ALMA primary beams.
We thank Shin'ichiro Asayama, Ted Huang, Hiroshi Nagai, Dirk Petry, George Moellenbrock and Charles Hull for providing clarification on the ALMA feed orientation.
\end{acknowledgements}

%
%

\end{document}